\documentclass[aps,showpacs,twocolumn]{revtex4}
\usepackage{amsfonts}
\usepackage{amsmath}
\usepackage{amssymb}
\usepackage{graphicx}
\usepackage{bm}

\input{tcilatex}
\begin{document}

\title{Trapping energy of a spherical particle on a curved liquid interface}
\author{Joseph L\'{e}andri and Alois W\"{u}rger}
\affiliation{LOMA, Universit\'{e} de Bordeaux \& CNRS, 351 cours de la Lib\'{e}ration,
33405 Talence, France}

\begin{abstract}
We derive the trapping energy of a colloidal particle at a liquid interface
with contact angle $\theta $ and principal curvatures $c_{1}$ and $c_{2}$.
The boundary conditions at the particle surface are significantly simplified
by introducing the shift $\varepsilon $ of its vertical position. We discuss
the undulating contact line and the curvature-induced lateral forces for a
single particle and a pair of nearby particles. The single-particle trapping
energy is found to decrease with the square of both the total curvature $%
c_{1}+c_{2}$ and the anisotropy $c_{1}-c_{2}$.\ In the case of non-uniform
curvatures, the resulting lateral force pushes particles toward more
strongly curved regions.\newline
\end{abstract}

\maketitle

\section{Introduction}

Colloidal particles trapped at a liquid phase boundary are subject to
capillary forces which induce pattern formation and directed motion \cite%
{Pie80,Bow97,Cav11}, and contribute to stabilize Pickering emulsions and
particle aggregates \cite{Koo11,But11}. Such microstructures affect the
mechanical and flow behavior of liquid and gel phases \cite{Ave03}, which in
turn are relevant for material properties and biotechnological applications 
\cite{Zen06}. In many instances, the particles are trapped at curved liquid
interfaces; rather surprisingly, even for spherical particles the influence
of curvature on capillary forces is not fully understood at present.

At a flat interface, capillary phenomena arise from normal forces induced by
the particle's weight or charge, or from geometrical constraints due to its
shape \cite{Cha81,Kra94}. As a simple example, an oat grain floating on a
cup of milk is surrounded by a meniscus that results from the its weight and
buoyancy; the superposition of the dimples of nearby grains reduces the
surface energy and thus causes aggregation. Charged beads exert electric
stress on the interface. The meniscus overlap of nearby particles causes a
repulsive electrocapillary potential \cite{For04,Oet05}, whereas beyond the
superposition approximation, a significantly larger attractive term is found 
\cite{Wue05,Oet05a,Dan10}. In the absence of gravity and electric forces,
capillary phenomena still occur for non-spherical particles: A capillary
quadrupole may arise from surface irregularities \cite{Sta00,Dan05}, pinning
of the contact line \cite{Fou02}, and for ellipsoids \cite%
{Nie05,Lou05,Mad09,Lew10}, and favors the formation of clusters with strong
orientational order.

A more complex situation occurs for interfaces with principal curvatures $%
c_{1}$ and $c_{2}$. The superposition of the weight-induced meniscus and the
intrinsic curvature results in a coupling energy that is linear in the
total\ curvature $H=c_{1}+c_{2}$. Its spatial variation gives rise to a
lateral force that drags a colloidal sphere along the curvature gradient 
\cite{Kra94a,Vas05}. Non-spherical particles interact through their
capillary quadrupole with the curvature difference $\delta c=c_{1}-c_{2}$,
and thus experience both a torque and lateral force \cite{Dom08}. The latter
is well known from the locomotion of meniscus-climbing insects and larvae,
which bend their body according to the local curvature such that the
capillary energy overcomes gravity \cite{Hu05,Bus06}; through a similar
effect, ellipsoidal particles prevent ring formation of drying coffee stains 
\cite{Yun11,Ver11}. A recent experiment on micro-rods trapped at a water-oil
meniscus illustrates both rotational and translational motion driven by
curvature \cite{Cav11}.


\begin{figure}[ptb]
\includegraphics[width=\columnwidth]{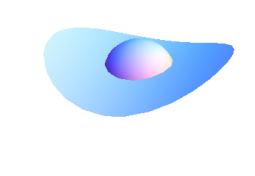}
\caption{Three-phase boundary of a spherical particle at a liquid interface with
curvatures $c_{2}=-\frac{1}{2}c_{1}$. The contact line is not a circle but
undulates in space. }
\label{Fig1}
\end{figure}

In this paper, we evaluate the geometrical part of the trapping energy of a
spherical particle on a curved interface; thus we consider only terms that
arise from the interface profile but are independent of body forces such as
weight and buoyancy. Previous papers considered limiting cases such as a
minimal surface ($H=0$) \cite{Wue06}, a spherical droplet ($\delta c=0$) 
\cite{Kra05,Kom06}, or a cylindrical interface ($H=\delta c$) \cite{Zen12};
yet a comprehensive picture is missing so far. Here we treat the general
case where both $H$ and $\delta c$ are finite, and obtain the trapping
energy in a controlled approximation to quadratic order in the curvature
parameters. We resort to the usual assumptions of constant contact angle $%
\theta $, curvature radius much larger than the particle size, and small
meniscus gradient.

As an original feature of the formal apparatus, we introduce the
curvature-induced shift $\varepsilon $ of the vertical particle position as
an adjustable parameter, in addition to the amplitude $\xi _{2}$ of the
quadrupolar interface deformation. As a main advantage, the boundary
conditions at the contact line separate in two independent equations for $%
\varepsilon $ and $\xi _{2}$, which are readily solved and provide a simple
physical picture for the effects of the two curvature parameters.

The paper is organized as follows. Section 2 gives a detailed derivation of
the energy functional and the deformation field $\xi (\mathbf{r})$. From the
usual variational procedure we find in section 3 the energy as a function of
the curvature parameters and the unkowns $\varepsilon $ and $\xi _{2}$; then
the energy is minimized with respect to the unknowns $\varepsilon $ and $\xi
_{2}$.\ In section 4 we show that the solution satisfies Young's law at the
three-phase boundary. In Section 5 we compare the trapping energy with
previous work, and discuss the contact line and curvature-induced forces.
Section 6 contains a brief summary.

\section{Trapping energy}

Here we derive the expression for the trapping energy and then evaluate it
explicitly to quadratic order in the curvatures. It consists of the surface
energies of all phase boundaries and the work done by the Laplace pressure
both on the liquid interface and on the area occupied by the particle.

First consider a particle dispersed in the liquid phase with the smaller
surface tension $\gamma _{m}=\min (\gamma _{1},\gamma _{2})$. The total
energy%
\begin{equation*}
\gamma S_{0}+W_{0}+\gamma _{m}4\pi a^{2}
\end{equation*}%
accounts for the interface area $S_{0}$, the work $W_{0}$, and for the
particle surface $4\pi a^{2}$, as illustrated in Fig. \ref{Fig3}a.

A particle approaching the interface gets trapped if the surface tensions
satisfy the inequality $|\gamma _{1}-\gamma _{2}|<\gamma $. The situation
shown in Fig. \ref{Fig3} corresponds to $\gamma _{m}=\gamma _{2}$. The total
energy%
\begin{equation*}
\gamma S+W+\gamma _{1}S_{1}+\gamma _{2}S_{2}
\end{equation*}%
consists of a term $\gamma S$ proportional to the area of the liquid
interface, the work $W$, and the particle segments in contact with the two
phases, $\gamma _{1}S_{1}+\gamma _{2}S_{2}$.

The trapping potential is given by the energy difference of these two
situations, 
\begin{equation}
E=\gamma (S-S_{0})+W-W_{0}+\gamma _{1}S_{1}+\gamma _{2}S_{2}-\gamma _{m}4\pi
a^{2}.  \label{6}
\end{equation}%
\ As illustrated in Fig. \ref{Fig3}b, $S$ is smaller than the unperturbed
area $S_{0}$. Since Young's law needs to be satisfied everywhere along the
three-phase contact line, $S$ may show a significantly more complex profile
than $S_{0}$.

In this section we evaluate the trapping energy to second order in the
curvature. There are two issues requiring particular care. First, both the
particle surface and the liquid interface contribute linear terms which,
however, cancel each other. Second, at quadratic order, there are various
contributions from the liquid interface, the area occupied by the particle,
and the work done by the Laplace pressure; these terms carry comparable
prefactors but opposite sign. The main result is given in Eq. (\ref{12})
below.

\subsection{Flat interface $H=0=\protect\delta c$}

\begin{figure}[ptb]
\includegraphics[width=\columnwidth]{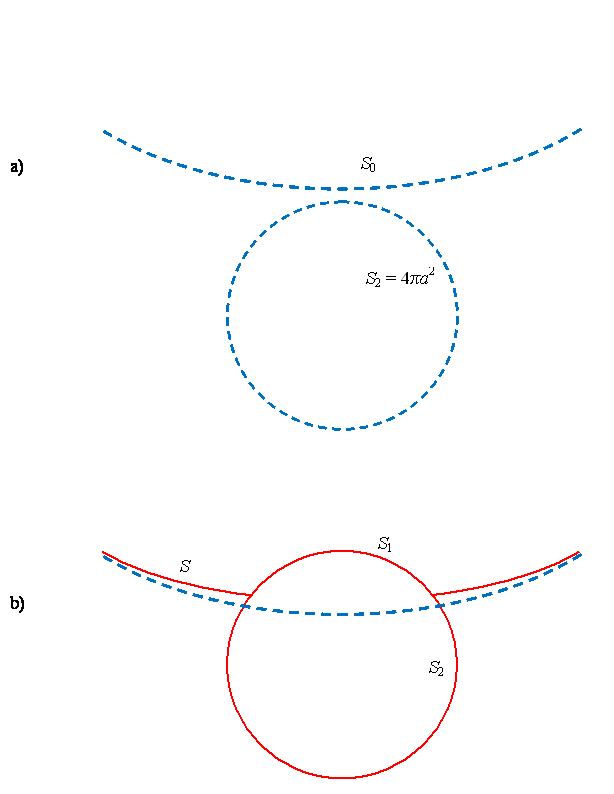}
\caption{Surface and interface areas
contributing to the trapping energy in Eq. (\protect\ref{6}). The upper
liquid is labelled \textquotedblleft 1\textquotedblright\ and the lower
\textquotedblleft 2\textquotedblright . a) The particle is in the phase of
lower surface energy (here $\protect\gamma _{2}<\protect\gamma _{1}$); the
liquid interface of area $S_{0}$ is described by (\protect\ref{1}). b)
Trapped state. The presence of the particle reduces the area of the liquid
phase boundary to the value $S$ and deforms its profile. The surface areas $%
S_{1}$ and $S_{2}$ are in contact with the two liquid phases. Note that the
figure shows one vertical section of the interface; both $S_{0}$ and $S$\
undulate when rotating about the vertical axis.}
\label{Fig2}
\end{figure}

We briefly recall the well-known results for zero curvature $w_{0}=0$, where
both $S_{0}$ and $S$ are\ flat \cite{Pie80}. Imposing local mechanical
equilibrium relates the surface tension parameters to the contact angle $%
\theta $ at the three-phase line in terms of Young's law 
\begin{equation}
\gamma _{1}-\gamma _{2}=\gamma \cos \theta .  \label{3}
\end{equation}%
Then the area of the liquid interface is reduced by 
\begin{equation*}
S-S_{0}=-\pi r_{0}^{2},
\end{equation*}%
and the segments of the particle surface read 
\begin{equation*}
S_{1}=2\pi a^{2}-2\pi az_{0},\ \ \ \ S_{2}=2\pi a^{2}+2\pi az_{0}.
\end{equation*}%
Here and in the following we use the vertical and radial coordinates of the
contact line, 
\begin{equation*}
z_{0}=a\cos \theta ,\ \ \ \ r_{0}=a\sin \theta ,
\end{equation*}%
as illustrated in the left panel of Fig.\ \ref{Fig2}. With Young's law one
finds for a flat interface \cite{Pie80}, 
\begin{equation}
E_{F}=-\pi a^{2}\gamma (1-|\cos \theta |)^{2}.
\end{equation}%
The trapping energy vanishes for contact angles $\theta =0$\ and $\theta
=\pi $. For $|\gamma _{1}-\gamma _{2}|>\gamma $ Young's law has no solution,
meaning that there is no stable trapped state. In the remainder of this
section we consider corrections to $E_{F}$ that arise at a curved interface.

\subsection{Curved interface without particles}

Now we consider the case of finite curvature. In Monge representation, $%
w_{0}(u,v)$ gives the interface height with respect to a tangent plane with
coordinates $u$\ and $v$.\ The energy consists of two terms, 
\begin{equation}
\gamma S_{0}+W_{0}=\gamma \dint dA\left( 1+\frac{1}{2}\left( \nabla
w_{0}\right) ^{2}\right) -\dint dAw_{0}P,  \label{1b}
\end{equation}%
the first of which describes the interface energy \cite{Wue06}, and the
second one the work done by the pressure difference $P$ between the two
sides of the interface. Here we have already used the small-gradient
approximation $|\nabla w_{0}|\ll 1$; for its derivation see \cite{Wue06}.
Its range of validity depends on the actual shape of the interface; for
experimentally relevant situations one finds that this approximation is
justified for distances wihtin the curvature radius. Since both the profile $%
w_{0}$ and the pressure $P$ turn out to be linear in $H$ and $\delta c$, Eq.
(\ref{1b}) is exact to second order in the curvature parameters.

The minumum-energy profile is determined by linearizing in terms of a small
fluctuation $\delta w_{0}$; integrating by parts one finds the corresponding
variation of energy 
\begin{equation*}
\delta E=-\int dA\ \delta w_{0}\left( \gamma \nabla ^{2}w_{0}+P\right) .
\end{equation*}%
Searching for a solution that is stable with respect to any small
deformation $\delta w_{0}$, we require $\delta E=0$ and thus find the
Young-Laplace equation 
\begin{equation}
\mathbf{\nabla }^{2}w_{0}+P/\gamma =0,  \label{1a}
\end{equation}%
which relates the profile to the pressure difference $P$ and the tension $%
\gamma $.

If the Laplace pressure varies sufficiently slowly along the interface, one
has $w_{0}=\frac{1}{2}\left( c_{1}u^{2}+c_{2}v^{2}\right) $, with the
coordinates $u$ and $v$ along the local principal curvature axes. For later
convenience we transform to polar coordinates; inserting $\ u=r\cos \varphi $%
\ and $v=r\sin \varphi $, resulting in 
\begin{equation}
w_{0}\left( r,\varphi \right) =\frac{r^{2}}{4}\left( H+\delta c\ \cos
(2\varphi )\right) .  \label{1}
\end{equation}%
The axes are chosen such that both $H$ and $\delta c$ are positive. Note
that the \textquotedblleft mean curvature\textquotedblright\ is often
defined as $H^{\prime }=\frac{1}{2}(c_{1}+c_{2})$ and thus differs from our $%
H$ by a factor $\frac{1}{2}$.

The Young-Laplace equation relates the mean curvature to the pressure
according to 
\begin{equation}
P=-\gamma H,  \label{1c}
\end{equation}%
whereas an asymmetry $\delta c$ is usually imposed by appropriate boundary
conditions. Eq. (\ref{1c}) takes a particularly simple form on a sphere of
radius $R$, where the curvature $H=2/R$ is related to the excess pressure $%
2\gamma /R$ inside the droplet.

Many experiments proble spatial variations of the curvature parameters $H$
and $\delta c$, which occur on a scale that is at least of the order the
curvature radius%
\begin{equation*}
R=2/\sqrt{H^{2}+\delta c^{2}}.
\end{equation*}%
Thus\ the quadratic form (\ref{1}) provides a good approximation for the
interface profile at distances within the curvature radius.

\subsection{Curved interface with a trapped particle}

Now we add a colloidal particle to the interface (\ref{1}). Like previous
papers, the present work relies on the separation of length scales, assuming
that the characteristic length of the deformation induced by a colloidal
particle of size $a$, is much larger than that of the unperturbed interface, 
$R$. All approximate formulae of the present paper can be cast in the form
of a truncated series in powers of $a/R$, and the trapping energy derived
below is exact to quadratic order.

Because of the undulating contact line, Young's law cannot be satisfied
along the intersection of $w_{0}(r,\varphi )$ and the spherical bead, but
requires a modified interface profile 
\begin{equation}
w=w_{0}+\xi ,  \label{2}
\end{equation}%
which is the sum of the unperturbed $w_{0}$ and the deformation field $\xi $%
. The latter has to be chosen such that the total energy is minimum.

The deformation field $\xi $ affects the trapping energy in three respects:
First, it modifies the work done by the Laplace pressure, second, it results
in a more complex profile of the liquid interface $S$ and, third, it
modifies the contact line, that is, the common boundary of $S$, $S_{1}$, and 
$S_{2}$.

\begin{figure}[ptb]
\includegraphics[width=\columnwidth]{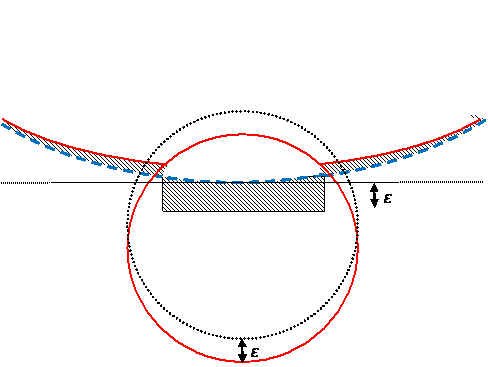}
\caption{ The hatched area gives
a schematic view of the volume of the work done by the Laplace pressure in
Eq. (\protect\ref{5a}). The volume corresponding to the interface domain $%
\mathcal{I}$ reads $\protect\int dA(w_{0}-w)$, and that over $\mathcal{P}$
reads $\protect\int dA(w_{0}+\protect\varepsilon )$, where $\protect%
\varepsilon $ is the vertical change of the particle position. The reference
state on a flat interface is shown as dotted lines. Solid and dashed lines
as in Fig. \protect\ref{Fig2}b.}
\label{Fig3}
\end{figure}

We start with the change in work due to the presence of a trapped particle.
It turns out convenient to separate the parameter space in domains $\mathcal{%
I}$ and $\mathcal{P}$, where $\mathcal{I}$ is the projection of the liquid
interface on the tangential plane and $\mathcal{P}$ the part occupied by the
particle. Then the work function reads 
\begin{equation}
W-W_{0}=\int_{\mathcal{I}}dA(w_{0}-w)P+\int_{\mathcal{P}}dA(w_{0}+%
\varepsilon )P.  \label{5a}
\end{equation}%
The first integral may be viewed as the change of potential energy of the
interface in the pressure field. The second one is proportional to the
vertical position of the particle with respect to the unperturbed interface.
In our notation, $\varepsilon >0$ corresponds to a downward motion of the
particle. In Fig.\ \ref{Fig3}he integration volume is shown as hatched area.

Now we turn to the modification of the liquid-interface area $S-S_{0}$.
Though the formally exact expressions can be given in terms of $\mathbf{%
\nabla }w$ and $\mathbf{\nabla }w_{0}$, \cite{Wue06}, we immediately use the
the small-gradient approximation and thus find 
\begin{eqnarray}
S-S_{0} &=&\frac{1}{2}\int_{\mathcal{I}}dA\left( (\mathbf{\nabla }%
w)^{2}-\left( \mathbf{\nabla }w_{0}\right) ^{2}\right)  \notag \\
&&-\int_{\mathcal{P}}dA\left( 1+\frac{1}{2}\left( \mathbf{\nabla }%
w_{0}\right) ^{2}\right) .  \label{5}
\end{eqnarray}%
The first integral accounts for the change of area of the deformed
interface, and the second one for the area occupied by the particle; the
corresponding parameter domains are $\mathcal{I}$ and $\mathcal{P}$.

Finally, we evaluate the change of the surface energy of the particle. The
segments in contact with the two liquid phases are given by the vertical
coordinate $\widetilde{z}(\varphi )$ of the contact line with respect to the
particle center, 
\begin{equation}
S_{1/2}=2\pi a^{2}\mp 2\pi a\int_{0}^{2\pi }d\varphi \widetilde{z}.
\label{11a}
\end{equation}%
Their sum obviously gives $4\pi a^{2}$.

\section{Energy minimization}

The trapping energy (\ref{6}) is a functional of the deformation field $\xi (%
\mathbf{r})$ and moreover depends on the vertical position parameter $%
\varepsilon $. In a first step we minimize $E[\xi ]$ with respect to the
shape function $\xi $, and thus obtain the trapping energy as a function of
the deformation amplitude and the vertical position. In a second step we
minimize with respect to the latter parameters, and thus obtain the energy
in terms of the curvature parameters.

\subsection{Deformation field}

Linearizing both $S-S_{0}$ and $W-W_{0}$ in terms of a fluctuation $\delta
\xi $ and integrating by parts, we find \cite{Wue06} 
\begin{equation*}
\delta E=-\int_{\mathcal{I}}dA\ \delta \xi \left( \gamma \mathbf{\nabla }%
^{2}\xi +\gamma \mathbf{\nabla }^{2}w_{0}+P\right) .
\end{equation*}%
The last two terms in parentheses cancel in view of Eq. (\ref{1a}). The
requirement that the energy be extremum, $\delta E=0$ for any $\delta \xi $,
directly leads to 
\begin{equation}
\mathbf{\nabla }^{2}\xi =0.  \label{4}
\end{equation}%
In other words, the deformation satisfies the equation of a minimal surface.

Integrating by parts and using (\ref{4}), we find for the first term in (\ref%
{5}) 
\begin{equation*}
-\int_{\mathcal{I}}dA\xi \mathbf{\nabla }^{2}w_{0}+\frac{1}{2}%
\doint\nolimits_{\partial \mathcal{I}}d\mathbf{s\cdot }\left( \mathbf{\nabla 
}\xi +2\mathbf{\nabla }w_{0}\right) \xi .
\end{equation*}%
From the Young-Laplace equation it is clear that the integral over $\mathcal{%
I}$ cancels the first term of the work (\ref{5a}). Inserting the remainder
in (\ref{6}) we obtain 
\begin{eqnarray}
E &=&\frac{1}{2}\doint\nolimits_{\partial \mathcal{I}}d\mathbf{s\cdot }%
\left( \mathbf{\nabla }\xi +2\mathbf{\nabla }w_{0}\right) \xi  \notag \\
&&-\int_{\mathcal{P}}dA\left( \gamma +\frac{\gamma }{2}\left( \mathbf{\nabla 
}w_{0}\right) ^{2}-(w_{0}+\varepsilon )P\right)  \notag \\
&&+\gamma _{1}S_{1}+\gamma _{2}S_{2}-\gamma _{m}4\pi a^{2}.  \label{7b}
\end{eqnarray}

\subsection{Truncation at second order}

Since the interface areas $S_{0}$ and $S$ are correct to second order in the
curvature, the different terms in the above energy are significant only to
quadratic order in the curvature and deformation parameters.

The general solution of (\ref{4}) reads $\xi _{0}\ln r+\sum_{k}\xi
_{k}\left( r{_{{0}}/r}\right) ^{k}\cos \left( k{\varphi }\right) $. The
logarithmic term vanishes in the absence of an external force such as
gravity; contributions with $k$ odd are absent for a spherical particle.
Because of the twofold symmetry of the source field (\ref{1}), the
quadrupolar term $k=2$ is the only contribution that is linear in the
curvature, the remaining coefficients are of least quadratic order. Thus we
write 
\begin{equation}
\xi (r,\varphi )=\xi _{2}\left( r{_{{0}}/r}\right) ^{2}\cos \left( 2{\varphi 
}\right)  \label{8}
\end{equation}%
and discard quadratic and higher-order terms in $\xi $. For notational
convenience we rewrite the unperturbed interface in the form 
\begin{equation*}
w_{0}(r,\varphi )=\frac{r^{2}}{r_{0}^{2}}\left( \omega _{0}+\omega _{2}\cos
(2\varphi )\right) ,
\end{equation*}%
with the parameters 
\begin{equation}
\omega _{0}=\frac{1}{4}Hr_{0}^{2},\ \ \ \ \ \omega _{2}=\frac{1}{4}\delta
cr_{0}^{2}.  \label{14}
\end{equation}%

\begin{figure}[t]
\includegraphics[width=\columnwidth]{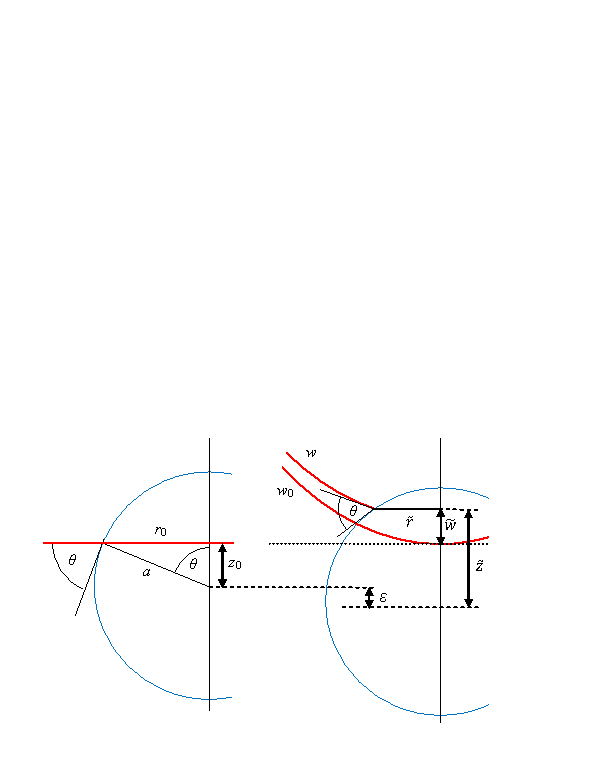}
\caption{ Side view on a sphere trapped
at a liquid phase boundary. The left panel shows a flat interface, where $%
r_{0}=a\sin \protect\theta $ and $z_{0}=a\cos \protect\theta $ are the
radial and vertical coordinates of the contact line with respect to the
particle center. The right panel illustrates the case of finite curvature.
In a vertical section of given azimuth, we show both the unperturbed
interface $w_{0}$ and the deformed profile $w$. Two phenomena concur in
order to satisfy Young's law: The interface profile changes by $\protect\xi %
=w-w_{0}$, and the particle adjusts its vertical position by $\protect%
\varepsilon $. Besides its radial coordinate $\tilde{r}$, we indicate the
vertical position of the contact line with respect to the tangent plane, $%
\tilde{w}$, and with respect to the particle center, $\tilde{z}=z_{0}+\tilde{%
w}+\protect\varepsilon $. In the case of \ finite curvature anisotropy $%
\protect\delta c$, the contact line is not a circle but undulates along the
particle surface, as illustrated by the top view in Fig. \protect\ref{Fig4}.}
\label{Fig4}
\end{figure}

\subsection{The contact line}

For further use we specify the radial and vertical coordinates $\widetilde{r}
$ and $\widetilde{z}$ of the undulating contact line. The curvature-induced
change of the vertical position with respect to that on a flat interface
comprises two terms,%
\begin{equation}
\widetilde{z}=z_{0}+\widetilde{w}+\varepsilon ,  \label{9}
\end{equation}%
where $\widetilde{w}=w_{0}(\widetilde{r})+\xi (\widetilde{r})$ accounts for
the vertical displacement of the contact line on the particles surface, and $%
\varepsilon $ for the change in the particle position with respect to the
tangential plane, as illustrated in the right panel of Fig. \ref{Fig2}. Any
point at the surface of the sphere satisfies the condition $\widetilde{r}%
^{2}+\widetilde{z}^{2}=a^{2}$, which can be rewritten as 
\begin{equation}
\widetilde{r}^{2}=r_{0}^{2}-2z_{0}\left( \widetilde{w}+\varepsilon \right) -(%
\widetilde{w}+\varepsilon )^{2}.  \label{10a}
\end{equation}

\subsection{Evaluation of $E$}

In the following we evaluate the trapping energy (\ref{7b}) to second order
in the curvature parameters $\omega _{0}$ and $\omega _{2}$, the deformation
amplitude $\xi _{2}$, and the vertical shift $\varepsilon $. Since the
integrand of the contour along $\partial \mathcal{I}$ is already of second
order, we take $d\mathbf{s}=-\mathbf{e}_{r}r_{0}d\varphi $, reduce the
gradients to the radial components $\partial _{r}\widetilde{\xi }+2\partial
_{r}\widetilde{w}_{0}$, and replace the coordinate $\widetilde{r}$ of the
contact line with $r_{0}$, 
\begin{equation*}
\frac{1}{2}\doint\nolimits_{\partial \mathcal{I}}d\mathbf{s\cdot }\left( 
\mathbf{\nabla }\xi +2\mathbf{\nabla }w_{0}\right) \xi =\pi \gamma \xi
_{2}^{2}-2\pi \gamma \xi _{2}\omega _{2}.
\end{equation*}%
The first term of the area integral is readily evaluated in terms of $dA=%
\frac{1}{2}\widetilde{r}^{2}d\varphi $; expanding $\widetilde{r}^{2}$ to
second order in the small parameters we find 
\begin{equation*}
\int_{\mathcal{P}}dA=\pi r_{0}^{2}-2\pi z_{0}(\omega _{0}+\varepsilon )-\pi
(\omega _{0}+\varepsilon )^{2}-\frac{\pi }{2}(\omega _{2}+\xi _{2})^{2}.
\end{equation*}%
In the following term the integrand $(\mathbf{\nabla }w_{0})^{2}$ is already
of second order in the curvature; thus we may replace the radius of the
contact line with $r_{0}$ and obtain 
\begin{equation*}
\frac{1}{2}\int_{\mathcal{P}}dA(\mathbf{\nabla }w_{0})^{2}=\pi \gamma \left(
\omega _{0}^{2}+\omega _{2}^{2}\right) .
\end{equation*}%
Noting $P=-\gamma H=-4\gamma \omega _{0}/r_{0}^{2}$, the work done on the
area occupied by the particle gives%
\begin{equation*}
\int_{\mathcal{P}}dA(w_{0}+\varepsilon )P=-2\pi \gamma \omega _{0}\left(
\omega _{0}+2\varepsilon \right) .
\end{equation*}%
Finally, we evaluate the change of the surface energy of the particle, 
\begin{equation*}
S_{1/2}=2\pi a^{2}\mp 2\pi a(z_{0}+\omega _{0}+\varepsilon ).
\end{equation*}

Inserting these expressions in (\ref{6}), separating the terms on a flat
interface, and replacing the surface tensions $\gamma _{1}$ and $\gamma _{2}$
through Young's law (\ref{3}), we find%
\begin{eqnarray}
E-E_{F} &=&\frac{3}{2}\pi \gamma \xi _{2}^{2}-\pi \gamma \xi _{2}\omega _{2}-%
\frac{\pi }{2}\gamma \omega _{2}^{2}  \notag \\
&&+\pi \gamma \varepsilon ^{2}-2\pi \gamma \omega _{0}\varepsilon -2\pi
\gamma \omega _{0}^{2}.  \label{12}
\end{eqnarray}%
Besides the curvature parameters $\omega _{0}$ and $\omega _{2}$, this
energy depends on two unknown parameters, the deformation amplitude $\xi
_{2} $ and the vertical shift $\varepsilon $ of the particle position with
respect to its value on a flat interface.

\subsection{Minimum energy}

The energy minimum is obtained from the zero of the derivatives with respect
to the adjustable parameters $\xi _{2}$ and $\varepsilon $, 
\begin{equation}
\frac{dE}{d\varepsilon }=0=\frac{dE}{d\xi _{2}}.  \label{7a}
\end{equation}%
From (\ref{12}) one readily finds the corresponding values

\begin{equation}
\varepsilon =\omega _{0},\ \ \ \ \xi _{2}=\omega _{2}/3.  \label{7}
\end{equation}%
In physical terms, the mean\ curvature $H$ results in a shift $\varepsilon $
of the particle toward the convex side of the interface. (In Fig. 3 this
means in downward direction.) On the other hand, the non-uniform curvature $%
\delta c$ gives rise to the quadrupolar amplitude $\xi _{2}$, which in turn
enhances the angular modulation of the interface.

Inserting the above values for $\varepsilon $ and\ $\xi _{2}$\ in the
trapping energy,\ we find%
\begin{equation}
E=E_{F}-3\pi \gamma \omega _{0}^{2}-\frac{2}{3}\pi \gamma \omega _{2}^{2}.
\label{15}
\end{equation}

\section{Young's law at the contact line}

The curvature-dependent part of $E$ has been calculated by minimizing the
total trapping energy with respect to the unknowns $\varepsilon $ and $\xi
_{2}$, without resorting to Young's law for the contact angle at the
three-phase boundary. Since Young's law is nothing else but the local
condition for a minimum-energy state, it expresses the same physical
constraint as Eq. (\ref{7a}) and thus provides an independent means of
checking the above results.\ 

This is achieved by imposing the contact angle $\theta $ along the
three-phase boundary. We start from the form \cite{Wue06}%
\begin{equation}
\cos \theta =\mathbf{n}_{I}\cdot \mathbf{n}_{P},  \label{13}
\end{equation}%
where $\mathbf{n}_{I}$ is the normal vector on the interface and$\ \mathbf{n}%
_{P}$ the normal on the particle surface. The former is best given in Monge
gauge with respect to the vertical axis, and the latter takes a simple form
because of the spherical geometry, 
\begin{equation}
\mathbf{n}_{I}=\frac{\mathbf{e}_{z}-\mathbf{\nabla }\widetilde{w}}{\sqrt{1+(%
\mathbf{\nabla }\widetilde{w})^{2}}},\ \ \ \ \mathbf{n}_{P}=\frac{\mathbf{e}%
_{z}\widetilde{z}+\mathbf{e}_{r}\widetilde{r}}{a}.  \label{13a}
\end{equation}%
Inserting in (\ref{13}) and linearizing in $\widetilde{w}$ and $\varepsilon $%
, we obtain the condition 
\begin{equation}
\widetilde{w}+\varepsilon -r_{0}\partial _{r}\widetilde{w}=0  \label{13b}
\end{equation}%
along the contact line. At linear order in the curvatures, we may replace
the radius $\widetilde{r}$ of the contact line with $r_{0}$, and thus put $%
\widetilde{w}=w(r_{0})$. Then (\ref{13b}) reduces to the simple algebraic
equation $-w_{0}(r_{0})+\varepsilon +3\xi (r_{0})=0$. Inserting the explicit
expressions for $w_{0}$ and $\xi $ gives%
\begin{equation}
-\omega _{0}+\varepsilon +\left( 3\xi _{2}-\omega _{2}\right) \cos 2\varphi
=0.
\end{equation}%
Solving for $\varepsilon $ and $\xi _{2}$ results in the same result as
those obtained from the minimization of the trapping energy in (\ref{7}).\ 

\section{Discussion}

The main results of the present paper are given by Eqs. (\ref{12}) and (\ref%
{15}). Here we discuss their most important features and compare with the
results of previous work.

\subsection{Vertical particle position}

Properly imposing Young's law along a non-circular contact line is not an
easy matter. Eq. (\ref{13}) relates the contact angle to the essential
parameters, the slope of the interface, as expressed by the gradient $\nabla 
\widetilde{w}$, and the vertical position $\widetilde{z}$ of the contact
line on the particle. Previous authors mostly chose to cast this in a
geometrical relation for the angle $\alpha $ of inclination of the
interface, $\tan \alpha =\nabla \widetilde{w}$, in the frame attached to the
particle. We found it helpful to introduce the vertical shift $\varepsilon $
of the particle position with respect to the tangential plane.

This approach leads to a rather simple relation of the contact angle to the
interface deformation, in terms of (\ref{13a}) and \ (\ref{9}). The
resulting linearized differential equation (\ref{13b}) comprises two
constraints: The term varying with the azimuthal angle determines the
deformation amplitude $\xi _{2}$, whereas the constant provides the vertical
shift $\varepsilon $. The rather simple solution (\ref{7}) shows that the
vertical shift is determined by the mean\ curvature, and the deformation
amplitude by the anisotropy $\delta c$.

The vertical shift is readily confirmed for the example of a particle
trapped on a spherical droplet of radius $R$, where our result $\varepsilon =%
\frac{1}{2}r_{0}^{2}/R$ can be obtained from the geometrical relation for
the particle position \cite{Kra05,Kom06}. On a cylindrical interface of
radius $R$, we find a vertical shift $\varepsilon =\frac{1}{4}r_{0}^{2}/R$;
this does not agree with the discussion in Ref. \cite{Zen12}, where a much
weaker shift $\propto a^{4}/R^{3}$ was obtained.

\subsection{Laplace pressure}

The work done by the Laplace pressure turns to be essential. Like previous
papers we have evaluated curvature effects by taking the flat interface as
reference state. Fig. 3 shows the work in terms of the integrated volume.
The integral over $\mathcal{I}$\ in (\ref{5a}) is cancelled the term $\xi 
\mathbf{\nabla }^{2}w_{0}$ in the interface energy (\ref{5}); in physical
terms the sum of the cost in deformation energy and the gain in work
vanishes.

Yet the second term in (\ref{5a}), that is, the integral over the area
occupied by the particle, results in a large negative contribution $-6\pi
\gamma \omega _{0}^{2}$ to the trapping energy. The physical meaning of the
integrand in $P\int dA(w_{0}+\varepsilon )$ becomes clear by comparing to
the reference state: When switching on the curvature, the particle-free
interface is shifted by $w_{0}$ with respect to the flat interface; thus
adding a particle results in the work $-P\int dA(-w_{0})$. On the other hand
the trapped particle position is shifted by $\varepsilon $ in downward
direction, corresponding to the work $-P\int dA(-\varepsilon )$. Thus both
contributions concur to a curvature-induced enhancement of the trapping
energy.

\subsection{Trapping energy}

The curvature-induced correction (\ref{15}) of the trapping energy is a
quadratic function of the curvature parameters $H$ and $\delta c$,%
\begin{equation}
E=E_{F}-\pi \gamma r_{0}^{4}\left( \frac{3}{16}H^{2}+\frac{1}{24}\delta
c^{2}\right) .  \label{16}
\end{equation}%
The numerical coefficients $-\frac{3}{16}$ and $-\frac{1}{24}$\ result from
several positive and negative terms of comparable size in (\ref{12}). Thus
it is essential to carefully evaluate all quadratic contributions to (\ref{6}%
). An increaise of either the total\ curvature $H$ or the anisotropy $\delta
c$ lowers the energy and enhances trapping.

We compare our Eq. (\ref{16}) to the results of previous work. In an earlier
paper \cite{Wue06}, one of us considered the case of a minimal surface ($H=0$%
) and found, in the notation adopted here, $E-E_{F}=-\frac{1}{24}\pi \gamma
r_{0}^{4}\delta c^{2}$, which agrees with the second term in (\ref{16}). In
Ref. \cite{Wue06} the Laplace pressure and the surface energy of the
particle had been discarded from the beginning; the more general approach of
the present work confirms that this is justified for $H=0$.

Kralchevsky et al. \cite{Kra05}\ and Komura et al. \cite{Kom06} calculated
the interface energy of a particle trapped on a liquid droplet of radius $R$
and total\ curvature $H=2/R$. Expanding their result in powers of $a/R$ and
truncating at second order, we obtain $\frac{3}{16}\pi \gamma r_{0}^{4}H^{2}$%
, in agreement with our expression for the surface energies; adding moreover
the work (\ref{5a}) done by the Laplace pressure, $-\frac{3}{8}\pi \gamma
r_{0}^{4}H^{2}$, we recover the trapping energy (\ref{16}).

More recently, Zeng et al. \cite{Zen12} considered the case of a particle
trapped at a cylindrical interface of radius $R$, where $H=\delta c=1/R$,
and found $E-E_{F}=\gamma r_{0}^{4}(\frac{3}{16}\pi -0.5333)/R^{2}$; the
first term $\sim \frac{3}{16}$ agrees with our result for the interface
energy. Note that Zeng et al. do not take into account the Laplace pressure.

\begin{figure}[ptb]
\includegraphics[width=\columnwidth]{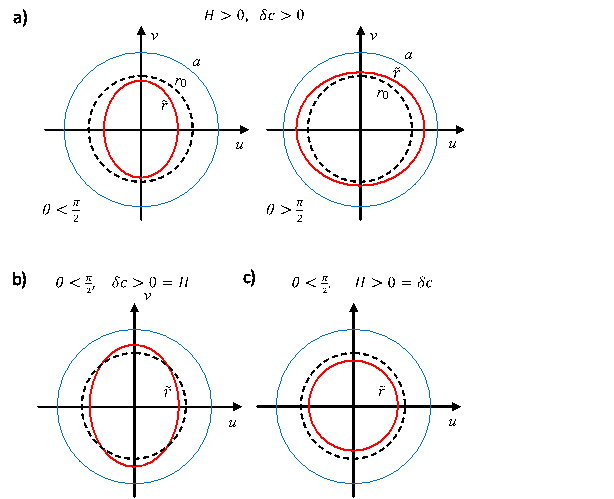}
\caption{ Top view of the three-phase
contact line on a particle of radius $a$. The dashed circle indicates the
contact line of radius $r_{0}$ at a flat interface. The solid (red) line
gives the radial coordinate $\widetilde{r}$ according to (\protect\ref{10}).
a) The upper panel shows the case where both curvature parameters $H$ and $%
\protect\delta c$ are positive and take similar values. For small contact
angles $\protect\theta <\frac{\protect\pi }{2}$, the radius $\widetilde{r}$
of the contact line is reduced according to (\protect\ref{10}), whereas it
increases for large contact angles $\protect\theta >\frac{\protect\pi }{2}$.
In both cases the effect is strongest along the axis $u$ with the largest
principal curvature $c_{1}$. The lower panel illustrates the case where
either $H$ are $\protect\delta c$\ vanish.\ b) For zero mean curvature $H=0$%
, that is on a minimal surface, the radial coordinate undulates about the
mean value $r_{0}$. c) In the case of zero asymmetry, $\protect\delta c=0$,
the curvature-induced change of the radial coordinate is constant, and $%
\widetilde{r}$ describes a circle.}
\label{Fig5}
\end{figure}

\subsection{Contact line}

It turns instructive to explicitly give the position of the contact line.
Inserting $\widetilde{w}$ and $\varepsilon $ in (\ref{9}) we have%
\begin{equation}
\widetilde{z}-z_{0}=\frac{r_{0}^{2}H}{2}+\frac{r_{0}^{2}\delta c}{3}\cos
(2\varphi ).  \label{10b}
\end{equation}%
The right-hand side is independent of the sign of $\cos \theta $. Thus in
the case of finite $H$, the contact line always moves toward the convex side
of the interface. If the anisotropy $\delta c$ exceeds the mean\ curvature,
the contact line may move in either direction on different parts of the
contact line, depending on the ratio $\delta c/H$.

Regarding the change of the radial position, we expand (\ref{10a}) to linear
order in $\widetilde{z}-z_{0}$ and find 
\begin{equation}
\widetilde{r}-r_{0}=-\frac{z_{0}}{r_{0}}\left( \frac{r_{0}^{2}H}{2}+\frac{%
r_{0}^{2}\delta c}{3}\cos (2\varphi )\right) .  \label{10}
\end{equation}%
Note that $r_{0}$ is always positive, whereas $z_{0}=a\cos \theta $ takes a
positive sign for small contact angles $\theta <\frac{\pi }{2}$, and a
negative one for $\theta >\frac{\pi }{2}$. Thus (\ref{10b}) and (\ref{10})
have opposite sign for small contact angles, and the same sign for large $%
\theta $.

The radial modulation, that is the projection of the contact line on the $%
u-v $-plane, is illustrated in Fig. \ref{Fig5}. The upper panel a) shows the
case of positive $H$ and finite $\delta c$, where the contact line moves
upward and undulates around the particle; for small contact angle, the
upward motion reduces the mean radius, whereas for $\theta >\frac{\pi }{2}$,
it is accompanied by an increase of the radius. Fig. \ref{Fig5}b) shows the
case of zero mean\ curvature and finite $\delta c$, where the radius
undulates along the contact line but its mean value is unchanged; a similar
picture occurs for $\theta >\frac{\pi }{2}$, albeit with the axes $u$ and $v$
exchanged. As a last example, Fig. \ref{Fig5}c) illustrates the case $\delta
c=0$, where the contact line remains a cercle.

Finally we note that the above expression for the contact line relies on the
quadrupolar approximation in Eq. (\ref{8}), which becomes exact at large
distances. Still, Eqs. (\ref{10b}) and (\ref{10}) provide a very good
description for the contact line as long as the radial change $\widetilde{r}%
-r_{0}$ can be linearized in terms of the meniscus deformation $\widetilde{z}%
-z_{0}$, in other words, as long as the derivative $d\widetilde{r}/d%
\widetilde{z}=-\cot \theta $ is finite. This implies that the quadrupolar
approximation (\ref{8}) ceases to be valid in the immediate vicinity of the
poles, $\theta \approx 0$ and $\theta \approx \pi $.

\subsection{Lateral force}

On an interface with spatially varying curvature, the trapping energy
changes with position and thus gives rise to a lateral force $\mathbf{F}=-%
\mathbf{\nabla }E$ on the trapped particle, 
\begin{equation}
\mathbf{F}=\pi \gamma r_{0}^{4}\left( \frac{3}{8}H\mathbf{\nabla }H+\frac{1}{%
12}\delta c\mathbf{\nabla }\delta c\right) .  \label{18}
\end{equation}%
As shown in our previous work \cite{Wue06}, the gradient of the curvature
anisotropy pushes the particle towards more strongly curved regions of the
interface. The numerical prefactor of the term proportional to $\mathbf{%
\nabla }H$ is larger by $\frac{9}{2}$.

\subsection{Two-particle interaction}

Finally we discuss curvature-induced forces between neighbor particles. Such
multipole interactions are well-known for non-spherical particles \cite%
{Sta00,Fou02,Dan05}; here we consider the mutual force on spheres trapped at
a curved interface. In a first step we derive the modified parameters $%
\widehat{H}$ and $\widehat{\delta c}$ that account for both the intrinsic
curvature and additional terms due to a colloidal particle. The parameter $%
\widehat{H}$ is given by $\nabla ^{2}w=\nabla ^{2}(w_{0}+\xi )$. Since the
deformation field $\xi $ obeys the equation (\ref{4}) of a minimal surface,
we find $\widehat{H}=H$; in other words, the particle does not change the
mean\ curvature of the interface. \ 

The anisotropy is best calculated in cartesian coordinates $u$ and $v$, where%
\begin{equation*}
\widehat{\delta c}=\delta c+\partial _{u}^{2}\xi -\partial _{v}^{2}\xi .
\end{equation*}%
This form is readily evaluated and gives after transformation to polar
coordinates 
\begin{equation}
\widehat{\delta c}=\delta c\left( 1+\frac{r_{0}^{4}}{r^{4}}\cos (4\varphi
)\right) .
\end{equation}%
Thus the deformation field $\xi $ significantly modifies the curvature in
the vicintiy of the particle.\ The additonal term decays with the fourth
power of the distance; because of its fourfold symmetry, the angular
modulation is maximum along the principal axes $u$ and $v$, and minimum in
between.

Each particle feels the additional curvature induced by its neighbor.
Superposition of their deformation fields gives the pair potential \cite%
{Wue06} 
\begin{equation*}
U=-\frac{\pi \gamma \delta c^{2}r_{0}^{8}}{48\rho ^{4}}\cos (4\varphi ),
\end{equation*}%
where\ $\rho $ and $\varphi $ describe the relative position of the
particles.\ With the corresponding unit vectors $\mathbf{e}_{\rho }$ and $%
\mathbf{e}_{\varphi }$ the mutual force reads as 
\begin{equation}
\mathbf{F}_{2}=-\frac{\pi \gamma r_{0}^{8}\delta c^{2}}{12\rho ^{5}}\left(
\cos (4\varphi )\mathbf{e}_{\rho }+\sin (4\varphi )\mathbf{e}_{\varphi
}\right) ,
\end{equation}%
Thus the capillary force between two nearby particle is not a central force:
Besides the attractive radial component, there is an additional force that
tends to align the particles parallel to one of the principal axes. As noted
previously, the latter force favors aggregates of cubic symmetry \cite{Wue06}%
.

A simple estimate shows that either of the forces $F_{2}$ and $F$ may
dominate. The curvature parameters vary on the scale of the curvature radius 
$R$, resulting in curvature-induced force $F\sim \gamma r_{0}^{4}R^{-3}$,
whereas that due to pair interactions decays on the scale of the particle
distance, $F_{2}\sim \gamma r_{0}^{8}\rho ^{-5}R^{-2}$. With typical values $%
r_{0}\sim 1$ $\mu $m and $R\sim 1$ mm, one finds that the ratio $F_{2}/F\sim
Rr_{0}^{4}\rho ^{-5}$ is larger than unity at distances of a few $r_{0}$,
and smaller than unity beyond.

\subsection{Comparison to gravity-curvature coupling}

So far we have discarded gravity effects. We conclude by comparing the
purely geometrical force (\ref{18}) with the well-known force arising for
heavy particles on a curved interface \cite{Cha81,Kra94a,Kra94,Vas05}. The
competition of weight and buoyancy results in an effective mass $m_{\text{eff%
}}=\frac{4}{3}\pi a^{3}\varrho _{\text{eff}}$, where 
\begin{align}
\varrho _{\text{eff}}& =\left( \varrho _{P}-\varrho _{u}\right) \left( \frac{%
1}{2}-\frac{3c_{0}}{4}+\frac{c_{0}^{3}}{4}\right)  \notag \\
& \ \ \ \ +\left( \varrho _{P}-\varrho _{l}\right) \left( \frac{1}{2}+\frac{%
3c_{0}}{4}-\frac{c_{0}^{3}}{4}\right)
\end{align}%
depends on the contact angle, $c_{0}=\cos \theta $, and on the densities of
the particle $\varrho _{P}$, and the upper and lower fluids, $\varrho _{u}$
and $\varrho _{l}$ \cite{Cha81}. $\varrho _{\text{eff}}$ may take either
sign, depending on the contact angle and on the density contrast of the
three phases. The meniscus around the particle is described by the
deformation field $\zeta (r)=-(m_{\text{eff}}g/2\pi \gamma )K_{0}(r/\ell )$,
where $K_{0}$ is a Bessel function and $\ell =\sqrt{\gamma /g\Delta \varrho }
$ the capillary length. Its coupling to the intrinsic curvature, 
\begin{equation}
E_{G}=\gamma \int dA\nabla \zeta \cdot \nabla w_{0},
\end{equation}%
is readily integrated, $E_{G}=m_{\text{eff}}gH\ell ^{2}$. A curvature
gradient leads to a lateral force that has been derived by several authors 
\cite{Cha81,Kra94a,Kra94,Vas05}; in our notation it reads 
\begin{equation}
F_{G}=-\gamma \pi a^{3}\frac{4}{3}\frac{\varrho _{\text{eff}}}{\Delta
\varrho }\nabla H.  \label{24}
\end{equation}%
where $\Delta \varrho $ is the density contrast of the fluids.

Comparison with Eq. (\ref{18})\ reveals that its first term $\sim \gamma \pi
a^{4}H\nabla H$ is by a factor $aH$ smaller than the weight-induced force $%
F_{G}$.\ This means that the geometrical force studied here is relevant if
(i) the first term $H\nabla H$ is significantly smaller than $\delta c\nabla
\delta c$, or if (ii) the gravity-induced forces are small, that is, if the
effective density $\varrho _{\text{eff}}$ is small as compared to the
density contrast $\Delta \varrho $ of the two fluids.

\section{Summary}

Starting from the well-known form (\ref{6}), we have evaluated the trapping
energy of a spherical particle to quadratic order in the curvature
parameters.

(i) On a formal level, the introduction of the shift $\varepsilon $ of the
vertical partical position, leads to a remarkably simple equation (\ref{13b}%
) for Young's law at the three-phase boundary, which is solved in (\ref{7}).
This could be useful for disentangling the involved boundary condtions
occurring at the surface of cylinders and ellipsoids \cite{Bot12}, or on
Janus particles \cite{Par12}.

(ii) An important contribution to the trapping energy results from the work
done by the Laplace pressure on the area occupied by the particle, that is,
from the second integral in (\ref{5a}).

(iii) As a main result, Eq. (\ref{16}) shows that both the total curvature $%
H $ and the anisotropy $\delta c$ lower the energy and thus enhance
trapping. As a consequence, both terms of the curvature-induced lateral
force (\ref{18}) drive particles toward strongly curved regions.

(iv) Eqs. (\ref{10b}) and (\ref{10}) give the radial and vertical
coordinates of the undulating contact line in terms of contact angle and
curvature parameters; the main dependencies are illustrated in Fig. \ref%
{Fig5}.

(v) The particle-induced interface deformation $\xi $ is proportional \ to
the curvature anisotropy $\delta c$ but independent of $H$. As a
consequence,\ only the anisotropy gives rise to a capillary interactions of
nearby particles, and the interaction potential $U$ reduces to the form
derived previously for $\delta c\neq 0=H$.


\begin{thebibliography}{99}
\bibitem{Pie80} P. Pieranski, Phys. Rev. Lett., \textbf{45}, 569 (1980).

\bibitem{Bow97} N. Bowden, A. Terfort, J. Carbeck, G.M. Whitesides, Science 
\textbf{11}, 233 (1997)

\bibitem{Cav11} M. Cavallaro, L. Botto, E.P. Lewandowski, M. Wang, K.J.
Stebe, PNAS \textbf{108}, 20923 (2011)

\bibitem{Koo11} E.\ Koos, N.\ Willenbacher, Science \textbf{311}, 897 (2011)

\bibitem{But11} H.-J.\ Butt, Science \textbf{311}, 868 (2011)

\bibitem{Ave03} R. Aveyard, B.P. Binks, J. H. Clint, Adv. Colloid Interface
Sci. \textbf{100--102}, 503 (2003)

\bibitem{Zen06} C. Zeng, H. Bissig, A.D. Dinsmore, Solid State Comm. \textbf{%
139}, 547, (2006)

\bibitem{Cha81} D.Y.C. Chan, J.D.\ Henry, L.R.\ White, J. Colloid Interf.
Sci. \textbf{79}, 410 (1981)

\bibitem{Kra94} P.A. Kralchevsky, K. Nagayama, Langmuir \textbf{10}, 23
(1994)

\bibitem{For04} L. Foret, A. W\"{u}rger, Phys. Rev. Lett.\textit{\ }\textbf{%
92}, 058302 (2004)

\bibitem{Oet05} M.\ Oettel, A.\ Dominguez, S.\ Dietrich, Phys. Rev E \textbf{%
71}, 051401 (2005)

\bibitem{Wue05} A. W\"{u}rger, L. Foret, J. Phys. Chem. B \textbf{109},
16435 (2005)

\bibitem{Oet05a} M.\ Oettel, A.\ Dominguez, S.\ Dietrich, J. Phys. Condens.\
Matter \textbf{17}, L337 (2005)

\bibitem{Dan10} K.D.\ Danov, P.A.\ Kralchevsky, J.\ Colloid Interf. Sci. 
\textbf{345}, 505 (2010)

\bibitem{Sta00} D.\ Stamou, D.\ Duschl, D.\ Johannsmann, Phys. Rev. E 
\textbf{62, }5263 (2000)

\bibitem{Dan05} K.D. Danov, P.A. Kralchevsky, B.N. Naydenov, G. Brenn J.\
Colloid Interf. Sci. \textbf{287}, 121 (2005)

\bibitem{Fou02} J.-B.\ Fournier, P. Galatola, Phys.\ Rev. E \ \textbf{65},
031601 (2002)

\bibitem{Nie05} E.A.\ van Nierop, M.A.\ Stejnman, S.\ Hilgenfeldt,
Euorphys.\ Lett. \textbf{72}, 671 (2005)

\bibitem{Lou05} J.C.\ Loudet, A.M. Alsayed, J. Zhang, A. G. Yodh, Phys.\
Rev.\ Lett. \textbf{94}, 018301 (2005)

\bibitem{Mad09} B. Madivala, J. Fransaer, J. Vermant, Langmuir \textbf{25},
2718 (2009)

\bibitem{Lew10} E.P. Lewandowski, et al., Langmuir \textbf{26}, 15142 (2010)

\bibitem{Kra94a} P.A. Kralchevsky, V.N.\ Paunov, N.D.\ Denkov, K. Nagayama,
J. Colloid Interf. Sci. \textbf{167}, 47 (1994)

\bibitem{Vas05} N.D. Vassileva, D. van den Ende, F. Mugele, J. Mellema,
Langmuir \textbf{21}, 11190 (2005)

\bibitem{Dom08} A.\ Dominguez, M.\ Oettel, S.\ Dietrich, J.\ Chem.\ Phys. 
\textbf{128}, 114904 (2008)

\bibitem{Hu05} D.L.\ Hu, J.W.M.\ Bush, Nature \textbf{437}, 733 (2005)

\bibitem{Bus06} J.W.M. Bush, D.L.\ Hu, Ann. Rev. Fluid Mech. \textbf{38},
339 (2006)

\bibitem{Yun11} P.J. Yunker, T. Still, M.A. Lohr, A.G. Yodh, Nature \textbf{%
476}, 308 (2011)

\bibitem{Ver11} J. Vermant, Nature \textbf{476}, 286 (2011)

\bibitem{Wue06} A. W\"{u}rger, Phys. Rev.\ E \textbf{74}, 041402 (2006)

\bibitem{Kra05} P.A. Kralchevsky, I.B. Ivanov, K.P. Ananthapadmanabhan, A.
Lips, Langmuir \textbf{21}, 50 (2005)

\bibitem{Kom06} S. Komura, Y. Hirose, Y. Nonomura, J.\ Chem.\ Phys. \textbf{%
124}, 241104 (2006)

\bibitem{Zen12} C. Zeng, F. Brau, B. Davidovitcha, A.D. Dinsmore, Soft
Matter \textbf{8}, 8582 (2012)

\bibitem{Bot12} L.\ Botto, L.\ Yao, R.L.\ Leheny, K.J.\ Stebe, Soft Matter 
\textbf{8}, 4971 (2012)

\bibitem{Par12} B.J.\ Park, D.\ Lee, ACS\ Nano \textbf{6}, 782 (2012)
\end{thebibliography}
\end{document}